%% file: main.tex
\documentclass[sigconf]{acmart}

\setcopyright{acmlicensed}
\copyrightyear{2025}
\acmYear{2025}
\acmDOI{XXXXXXX.XXXXXXX}
\acmConference[arXiv '25]{arXiv Preprint}{July 05,
  2025}{Woodstock, NY}

\usepackage{subcaption}

\usepackage[listings]{tcolorbox}
\usepackage{tcolorbox}

\newcommand{\dcc}{DCC}
\newcommand{\dcchelp}{\dcc{}~Help}
\newcommand{\gpt}{\textit{gpt-4.1-2025-04-14}}

\begin{document}


\title{Narrowing the Gap: Supervised Fine-Tuning of Open-Source LLMs as a Viable Alternative to Proprietary Models for Pedagogical Tools}




\author{Lorenzo Lee Solano}
\affiliation{%
  \institution{University of New South Wales}
  \city{Sydney}
  \country{Australia}}
\email{l.leesolano@unsw.edu.au}
\orcid{0009-0002-6079-1265}

\author{Charles Koutcheme}
\affiliation{%
  \institution{Aalto University}
  \city{Espoo}
  \country{Finland}}
\email{charles.koutcheme@aalto.fi}
\orcid{0000-0002-2272-2763}

\author{Juho Leinonen}
\affiliation{%
  \institution{Aalto University}
  \city{Espoo}
  \country{Finland}}
\email{juho.2.leinonen@aalto.fi}
\orcid{0000-0001-6829-9449}

\author{Alexandra Vassar}
\affiliation{%
  \institution{University of New South Wales}
  \city{Sydney}
  \country{Australia}}
\email{a.vassar@unsw.edu.au}
\orcid{0000-0001-8856-2566}

\author{Jake Renzella}
\affiliation{%
  \institution{University of New South Wales}
  \city{Sydney}
  \country{Australia}}
\email{jake.renzella@unsw.edu.au}
\orcid{0000-0002-9587-1196}

\renewcommand{\shortauthors}{Lee Solano et al.}

\begin{abstract}
Frontier Large language models (LLMs) like ChatGPT and Gemini can decipher cryptic compiler errors for novice programmers, but their computational scale, cost, and tendency to over-assist make them problematic for widespread pedagogical adoption. This work demonstrates that smaller, specialised language models, enhanced via Supervised Fine-Tuning (SFT), present a more viable alternative for educational tools. We utilise a new dataset of 40,000 C compiler error explanations, derived from real introductory programming (CS1/2) student-generated programming errors, which we used to fine-tune three open-source models: Qwen3-4B, Llama-3.1-8B, and Qwen3-32B. We performed a dual evaluation, combining expert human reviews with a large-scale automated analysis of 8,000 responses using a validated LLM-as-judge ensemble. Our results show that SFT significantly boosts the pedagogical quality of smaller models, achieving performance comparable to much larger models. We analyse the trade-offs between model size and quality, confirming that fine-tuning compact, efficient models on high-quality, domain-specific data is a potent strategy for creating specialised models to drive educational tools. We provide a replicable methodology to foster broader access to generative AI capabilities in educational contexts.
\end{abstract}  





\maketitle

\input{sections/01_introduction}
\input{sections/02_related_work}

\input{sections/03_methods}
\input{sections/04_results}

\section{Discussion}
\input{sections/05_discussion}

\section{Conclusion}
\input{sections/06_conclusion}

\input{sections/07_acknowledgements}

\balance
\bibliographystyle{ACM-Reference-Format}
\bibliography{ref,references_SV,references}



\end{document}

%% file: sections/01_introduction.tex
\section{Introduction}
Large language models (LLMs) are increasingly being integrated into computing education~\cite{prather2025beyond,Denny2024ComputingAI} where they are used for a variety of tasks such as generating instructional content~\cite{Sarsa2022AutomaticModels,Logacheva2024Evaluating95-113.,tran2023generating}, generating automated feedback on buggy code~\cite{Hellas2023ExploringRequests,koutcheme_2024_open,Phung2023GeneratingModels,kiesler2023exploring,Pankiewicz2023LargeAssignments,ashok-kumar-lan-2024-improving}, code explanations~\cite{MacNeil2022ExperiencesE-Book,Leinonen2023ComparingModels,Sarsa2022AutomaticModels}, and tutoring~\cite{Liffiton2023CodeHelp:Classes}. This growing interest is also visible in the number of studies published focusing on generative AI in computing education~\cite{prather2025beyond}. One of the earliest and most prominent use cases has been helping students to understand programming error messages~\cite{Leinonen2023UsingMessages, Taylor2024DccModels,Wang2024ACS1,Santos2023AlwaysEnhancement,Santos2024NotPractice}. 

Existing approaches have largely relied on proprietary LLMs accessed through third-party APIs. While these models are powerful and seem to be helpful, this approach brings a number of practical and ethical limitations. Student code and context must be transmitted to external servers, raising data privacy concerns, particularly in educational environments \cite{Huang2023EthicsProtection}. In addition, the use of commercial APIs for accessing LLMs often comes with significant financial costs and operational risks, including rate limits, service outages, and changes in model behaviour over time.

Addressing these challenges, we examine the effectiveness of smaller open-source language models, for generating error message explanations in introductory programming (CS1/2).  While less capable out-of-the-box \cite{koutcheme2025evaluating}, these models can be fine-tuned to perform well on specific tasks \cite{Kotalwar2024Hints-In-Browser:Generation}. Our research question in this work is as follows:


 \begin{itemize}
    \item[\textbf{RQ}] How does Supervised Fine-Tuning on a dataset of CS1/2 error explanations affect the performance of smaller open-source models?
\end{itemize}

To investigate this, we fine-tuned a range of open-source language models using a large dataset of errors encountered by students and explanations generated by GPT-4.1, which was one of the top proprietary models available at the time of the study. We then compare the small fine-tuned models to proprietary LLMs to evaluate their explanations with regard to clarity and correctness, using a combination of expert and LLM-based annotation techniques for evaluation. 
Our contributions are threefold:
\begin{itemize}
    \item We provide a comprehensive benchmark of fine-tuned open-source models compared to commercial LLMs for the task of explaining programming error messages.
    \item We introduce a rubric for assessing the quality of programming error message explanations.
    \item We employ both expert judges and LLM-as-judge, with inter-rater reliability between the two groups to scale evaluation to a large dataset. 
\end{itemize}

%% file: sections/02_related_work.tex
\section{Related Work}

\paragraph{\textbf{LLMs for Programming Error Messages.}}
%
Research has begun to explore how LLMs can improve programming error messages (PEMs). 
For example, Leinonen et al.~\cite{Leinonen2023UsingMessages} found that Codex, a 2021 coding model, could enhance error messages about half of the time. Taylor et al.~\cite{Taylor2024DccModels} found that GPT-3.5 could explain errors in 90\% of the cases for compile-time and 75\% of the cases for run-time errors. Santos et al.~\cite{Santos2023AlwaysEnhancement} found that providing code context to the LLM improved the generated error message explanations. Wang et al.~\cite{Wang2024ACS1} found that students using LLM-enhanced error messages repeated fewer errors, and resolved errors faster. In all cases, models display a propensity to overhelp, which contradicts pedagogical goals \cite{Renzella2025Compiler-IntegratedPrograms, Liffiton2023CodeHelp:Classes, Hellas2023ExploringRequests}. More recent work evaluating LLM-enhanced compiler error messages from the student perspective has found them to be ineffective in practice \cite{Santos2024NotPractice}, largely due to the structure and detail included in the error message, since longer error explanations have previously been shown to be unhelpful~\cite{Nienaltowski2008CompilerNovices}.


\paragraph{\textbf{Fine-tuning LLMs in CS Education.}}
Few works in programming education have attempted to fine-tune open-source language models, despite privacy and reliability advantages. Hicke et al. \cite{hicke2023aita} developed and evaluated an LLM-based solution for answering student questions on a programming forum. The authors trained medium size LLaMa-2 models and compared them against alternatives. However, the evaluation was constrained by a small sample size of 20 responses rated by a single annotator, highlighting a need for larger-scale human evaluations with multiple annotators to ensure robust inter-rater reliability.
Kumar et al. \cite{ashok-kumar-lan-2024-improving} trained a Llama-3.1-8B model to provide guidance to students through asking Socratic questions. Their approach combined SFT and reinforcement learning, leveraging existing human annotations combined with LLM-generated alternatives. While this approach showcases the effectiveness of training open-source models, the method relies on costly human annotations, and does not support students with programming error messages.
Kotalwar et al. \cite{Kotalwar2024Hints-In-Browser:Generation} trained a Phi-mini model and Llama-3.1-8B model to produce in-browser feedback and Socratic hints for student code. Their results suggested that fine-tuned smaller models could attain performance parity with GPT-3.5 across three Python problem sets \cite{Kotalwar2024Hints-In-Browser:Generation}.

The work presented in this paper extends prior research and lessons learned, with the following core novelties: a) it benchmarks a variety of open-source models across parameter sizes, b) it is tightly integrated into a C compiler, which extends the data collection context to provide better explanations, and c) it provides a distinct evaluation of both compile- and run-time errors.

%% file: sections/03_methods.tex
\section{Methods}

Our goal is to support novice C programmers by providing guidance for both compile- and run-time errors. While existing solutions rely on proprietary LLMs \cite{Wang2024ACS1, Taylor2024DccModels, Liffiton2023CodeHelp:Classes, Liu2025ImprovingLearning},
we focus on the development of fine-tuned open-source alternatives, specialised for delivering pedagogically sound error message explanations.
More specifically, we rely on Supervised Fine-Tuning (SFT), which offers a cost-effective approach to refining a language model for specialised or domain-specific tasks.

\subsection{Data}
Training and evaluation data was sourced from student use of \dcchelp{}, a compiler-integrated approach to generating error message explanations for novice C programming students.
For five teaching periods at a large Australian university, between September 2023 and February 2025, we collected approximately 180,000 compile-time and 50,000 run-time student invocations of \dcchelp{}.
Each logged example includes the source code, the context of the error, as well as the original \dcchelp{} response generated by either GPT-3.5 Turbo or GPT-4o mini, which we use as a baseline of model performance.

All logged data is pre-processed and redacted in a best effort to remove identifying features, such as student IDs, names and emails, in line with the relevant ethics requirements. We also filtered out a few instances of unbounded recursion, which produced logs of excessive length, unsuitable for training or inference.

\subsection{Creating a Training Dataset}

Producing a representative labelled dataset required for SFT, consisting of example inputs and desired responses, can be challenging for the application of explaining error messages. 
As such, past work often utilised small curated or handcrafted datasets \cite{Salmon2025DebuggingEffectiveness, Liu2025ImprovingLearning}. 
To address this, we propose to distil the abilities of a large language model into smaller models \cite{Kotalwar2024Hints-In-Browser:Generation}. We generate training data by using a proprietary LLM to provide responses for examples in the \dcchelp{} dataset, resulting in a larger and more diverse training dataset. 

\paragraph{\textbf{Data Processing.}}
We sample 40,000 examples from the first two teaching sessions in the \dcchelp{} dataset, reserving the remaining sessions for evaluation.
As invocations of \dcchelp{} are not evenly distributed over time, a purely random sample risks disproportionately representing certain weeks and types of examples, such as those common near the deadlines of assignments. 
Instead, before taking a random sample, we first reduced the data pool by limiting the number of compile- and run-time examples from any given in-session week to a maximum (4,500 for compile-, 2,250 for run-time).

\paragraph{\textbf{Prompting Strategy.}} 
\label{para:prompt-strategy}
We build on the approach of Taylor et al.~\cite{Taylor2024DccModels} to construct prompts based on the available error context, such as compiler messages, source code, call-stack, and variable states. Our revised strategy, shown in Figure~\ref{fig:generate_feedback}, additionally enforces a consistent three-part explanation structure that draws from two distinct strands of prior work~\cite{Phung2023GeneratingModels, Kotalwar2024Hints-In-Browser:Generation}, synthesising research on error message interpretation with methods for structured feedback generation.

\input{figures/generation_prompt}

The first step asks the model to clarify the error message in simple, accessible language. This brief, jargon-free, natural language \textit{translation} of the original compiler message addresses common readability issues present in standard PEMs~\cite{Denny2021OnFactors}. 
The next two steps are motivated by the literature on programming feedback generation~\cite{Phung2023GeneratingModels}: we prompt the model to explain potential causes for the error, then provide actionable hints and guidance to help the student fix their code. This mirrors the approach of Phung et al.~\cite{Phung2023GeneratingModels,Kotalwar2024Hints-In-Browser:Generation}, using explanations of student errors as a chain-of-thought scaffold for generating higher-quality hints. 

\paragraph{\textbf{Dataset Generation.}}
Using this prompt strategy, we generate responses for the 40,000 examples using \gpt{}, via OpenAI's batch completions API, with a temperature of 0.

The resulting SFT training dataset consists of a ratio of approximately 3:1 compile- vs run-time examples, with total lengths ranging from 300 to 4,000 tokens (mean length of 849 tokens).

\subsection{Fine-tuning Open-Source Models}

Using the resulting training dataset, we train three open-source language models of various sizes: Qwen3-32B \cite{yang2025qwen3technicalreport}, Llama-3.1-8B \cite{grattafiori2024llama3herdmodels}, and Qwen3-4B \cite{yang2025qwen3technicalreport}. 
We chose these models to have a representative sample of different sizes. While Qwen3-32B is not a smaller language model, it allows us to contextualize the efficacy of our training approach.

\paragraph{\textbf{Fine-tuning Details.}}
All models were trained for 1 epoch, with a learning rate of $2e-5$. Training was performed through the Unsloth fine-tuning API, on a single Nvidia A100 GPU (80GB vram). 

We train all our models using QLoRa \cite{dettmers2023qloraefficientfinetuningquantized}, a Parameter-Efficient Fine-Tuning technique (PEFT) that quantises a language model to 4-bit and adds on top a set of trainable parameters called adapters, while the base model remains frozen. Using PEFT techniques has two benefits: quantized models allow easy edge device deployment, while adapters allow easy recovery of the original functionalities. 


\subsection{Evaluation Dataset} \label{sec:evaluation-dataset}

To evaluate the performance of the fine-tuned open-source models, we construct an evaluation dataset of 8,000 examples, sourced from the remaining three teaching periods in the \dcchelp{} dataset (from February 2024 to February 2025). Similar to our training dataset construction, we first limited the number of examples per week to a maximum (3,000 for compile-, 1,500 for run-time), before randomly sampling.

For this subset, we generate responses using the previously listed base and fine-tuned open-source models, using the Unsloth inference API, on a single A100 GPU. For comparison, we include the original \dcchelp{} response and a proprietary LLM response generated by \gpt{}. All responses were generated using the same prompt strategy, described in Section~\ref{para:prompt-strategy}. 
This produced an evaluation set of approximately 5,600 compile- and 2,400 run-time examples, each of which contains the student code and error context, and eight model responses (total evaluation responses: 64,000). 
From this, we randomly sampled a smaller dataset of 50 run-time and 50 compile-time examples for the expert (human) evaluation.


\subsection{Evaluation Rubric} \label{sec:rubric}

We define an annotation rubric of eight binary criteria, grounded in prior work on evaluating programming feedback \cite{koutcheme2025evaluating, Taylor2024DccModels,Hellas2023ExploringRequests}.
Each criterion is scored as either 0 or 1, with 1 indicating that the explanation exhibits the desired property. 

\begin{itemize}
    \item \textsc{Correctness}: The explanation is technically correct.
    \item \textsc{Selectivity}: Contains no incorrect/irrelevant information.
    \item \textsc{Completeness}: Contains all information critical to understand the error.
    \item \textsc{Clarity}: Clear, easy to understand, presented in a readable format, using an economy of words.
    \item \textsc{Novice Appropriate}: Accessible for novices, avoiding technical jargon and advanced knowledge assumptions.
    \item \textsc{No Solution}: Does not directly provide the full solution, either in code or prose.
    \item \textsc{No Overhelp}: Avoids over-direction, leaving room for problem solving and critical thinking.
    \item \textsc{Socratic}: Provides guidance to solve the error, and includes at least one relevant guiding question or statement.
\end{itemize}

Criteria such as \textsc{Correctness} and \textsc{Selectivity} are common quality criteria for evaluating programming feedback \cite{koutcheme2025evaluating, Kotalwar2024Hints-In-Browser:Generation, Taylor2024DccModels},  
and address typical failure modes of LLM-generated explanations. 
\textsc{Clarity} and \textsc{Novice Appropriate} reflect established recommendations for effective feedback for novice programmers \cite{Denny2021OnFactors}. 
\textsc{No Overhelp} and \textsc{Socratic} are drawn from prior work by Ross et al. \cite{Ross2025SupervisedEducation}, and focus on the pedagogical function of these explanations in the context of introductory programming. Finally, we take a pedagogical interpretation of \textsc{Completeness}, focusing on whether all information needed to understand the error message is present.

These criteria align with our explanation prompting strategy (see Figure~\ref{fig:generate_feedback}), where \textit{Error Message Clarification} helps to address \textsc{Clarity}, \textsc{Novice Appropriate} and \textsc{Completeness}. \textit{Potential Causes}, while not explicitly mapping to any criteria, acts in part as a reasoning step to help address \textsc{Correctness} and \textsc{Selectivity}; and \textit{Guidance} aligns with pedagogically oriented criteria such as \textsc{Socratic} and \textsc{No Overhelp}. 
Overall, the structural constraints within the feedback prompt are designed to promote novice-appropriate guidance and prevent the inclusion of explicit solutions. 

\subsection{Expert Evaluation}

Expert evaluation was performed by four experienced CS1 teaching staff members, 
who each annotated 20 unique examples. An additional subset of 20 examples was annotated by all four experts, which was used to calculate inter-rater reliability (IRR) using Gwet’s AC1 \cite{Gwet2008ComputingAgreement}. This brings the number of annotations per expert to 40 examples or 320 responses.
Gwet's AC1 was chosen as some metrics had very high degrees of agreement between raters, which introduced errors with more typical IRR metrics. 




Annotators were presented with the source code and error context for each example and were tasked with first annotating each model response against the evaluation rubric before ranking the same model responses from best to worst (with ranking 1 being best, and ranking 8 being worst). 
The model associated with each response was not revealed to annotators, and the order of model responses was randomised.

\subsection{LLM Evaluation}

We adopt the LLM-as-judge paradigm to label the expanded evaluation dataset of 8,000 examples, comprising 32,000 model-generated responses~\cite{zheng2023judging}.
Prior work has demonstrated the viability of using automatic LLM-generated annotations to scale evaluations in the domain of pedagogical explanations and feedback \cite{koutcheme_2024_open,koutcheme2025evaluating, Scarlatos2024ImprovingLearning, hicke2023aita}.

\paragraph{\textbf{Ensemble of Judges.}}
Rather than relying on a single judge, we use a panel of three strong LLMs: GPT-4.1, Gemini-2.5-Flash, and Qwen3-32B. 
Verga et al.
\cite{verga2024replacingjudgesjuriesevaluating} demonstrate that ensembles composed of diverse model families can outperform individual large models, particularly by mitigating model-specific biases.

To annotate the expanded evaluation dataset, we prompt each of the three judges individually to provide binary decisions across all evaluation criteria.
We adopt the Single Answer Grading approach proposed by Koutcheme et al.~\cite{koutcheme2025evaluating}, which utilises an additional reasoning step to generate annotations. The judge first generates its own error explanation in the first conversation turn, then compares and evaluates the candidate model's explanation in the second turn. 
As a key modification, we remove the explanation structure constraints (described in Section~\ref{para:prompt-strategy}) from the conversation history prior to the second turn, as the judges would inadvertently apply it to their reasoning and annotations.

From the ensemble, we obtain the final verdict using a strict unanimity policy: a criterion is marked correct only if all judges agree. While this method does not provide absolute performance guarantees, as discussed in Section \ref{sec:limitations}, it offers a consistent, scalable, and reliable strategy for relative comparisons.

Annotations for Qwen3-32B were generated via the vLLM interface on a single A100 GPU, with the other two judges accessed through model specific proprietary APIs.
Outputs were produced with a temperature of 0, reasoning disabled, and all other inference parameters set to default.

We validate IRR between LLM-judged and expert annotations across each metric using Gwet's AC1.

%% file: figures/generation_prompt.tex
\begin{figure}[thbp]
\centering
\begin{tcolorbox}[colback=gray!5, colframe=black, title=Feedback Prompt Template, fonttitle=\bfseries, boxrule=0.5pt, arc=2mm, left=1mm, right=1mm, top=0.5mm, bottom=0.5mm,before skip=0pt, after skip=0pt]
\footnotesize

\textbf{System description:} You are providing programming error messages in an intro programming course using C.

\textbf{Your inputs are:} C Program, Original Error, Variables Stack, Call Stack. \\


\textbf{Output the following:}

\begin{enumerate}
    \item 
    \textbf{Error Message Clarification:} Write one short sentence, explaining the error message without programming jargon. \\
    
    \item \textbf{Potential Causes:} Follow with 1-2 short sentences, identifying and 
   explaining potential issues in the code that may be 
   causing this error. \\

    \item \textbf{Guidance (Hints Generation):} Follow with 1-2 short sentences, giving debugging hints and guidance, potentially including specific references to my code.
\end{enumerate}

\textbf{Details:}  
\begin{itemize}
    \item Keep your response 
short, friendly, and without jargon.
\item Do not give the solution in code.
\item Address the student directly.
\end{itemize}

\end{tcolorbox}


\caption{Three-Step Error Explanation Prompt Strategy.}
\Description[The prompt structure used to generated error explanations]{Fully described in the text.}
\label{fig:generate_feedback}
\end{figure}

%% file: sections/04_results.tex
\section{Results}

\subsection{Expert Response Rankings}
Table \ref{tab:winrate-compile-runtime} lists the proportion of responses where each SFT open-source model was ranked above the base and proprietary models.


\input{figures/rank_win_rates}

Both SFT-Llama and SFT-Qwen-4B are preferred over their untrained counterparts, in both compile- and run-time cases. For Qwen-32B, the untrained variant is ranked higher slightly more often.
SFT-Llama performs equally well on both compile-time (0.74) and run-time errors (0.72) compared to its base counterpart. In contrast, while the SFT-Qwen-4B is significantly preferred over its base counterpart for compile-time errors, it is only slightly favoured for run-time errors (about 58\% of cases).
All SFT models are also preferred to original \dcchelp{} responses, especially for generating run-time error explanations. 

\input{figures/rank_scores}

GPT-4.1 achieved the best mean rank score, followed by both variants of Qwen-32B. The SFT variants of both Llama and Qwen-4B demonstrated improved rank scores, first place and last place rates over their base counterparts. The base Qwen-4B, Llama and the \dcchelp{} responses consistently ranked low, with first place rates of 0.01, 0.02 and 0.06 respectively. 

\subsection{Expert and LLM-Judged Metrics}


\paragraph{\textbf{Understanding Results.}}
The results of both the expert annotations (800 model responses) and LLM-judged annotations (64,000 model responses) are presented side-by-side in Table~\ref{tab:merged-expert-llm-raw}.
For the subset with annotations from all four experts, only one random annotation per example was included in the aggregate calculations, to ensure equal weighting. 

\input{figures/llm_judge/combined_eval}

\paragraph{\textbf{Performance Comparisons.}}
GPT-4.1 showed the strongest performance across the majority of metrics, as evidenced by both expert and LLM-judged evaluations. Specifically, GPT-4.1 achieved the highest rates for \textsc{Correctness}, scoring 0.92 in expert and 0.97 in LLM-judged evaluation; \textsc{Selectivity}, scoring 0.90 in expert and 0.95 in LLM-judged evaluation; \textsc{Novice Appropriate}, scoring 0.89 in expert and 0.99 in LLM-judged evaluation; and \textsc{No Overhelp}, scoring 0.95 in expert and 0.97 in LLM-judged evaluation. 

Base and SFT variants of Qwen-32B exhibited performance competitive with GPT-4.1 across many metrics. They matched GPT-4.1's performance in \textsc{Correctness} and \textsc{No Solution}, with the fine-tuned variant actually outperforming when judged by experts. For the remaining criteria, both Qwen-32B variants performed within a 0.10 or better margin of GPT-4.1.

While both base and SFT variants of Qwen-32B perform strongly overall, their relative performance differed between annotation strategies. 
Base Qwen-32B generally exhibited higher performance across LLM-judged results, achieving higher rates in metrics such as \textsc{Completeness} (0.63 for base versus 0.38 for SFT), and \textsc{Socratic} (0.44 for base versus 0.36 for SFT). Additionally, the base Qwen-32B variant showed the highest performance of any model in LLM-judged \textsc{all} at both compile- (0.34) and run-time (0.26).
In contrast, SFT-Qwen-32B outperformed its base counterpart across several expert-judged metrics, including \textsc{No Solution} (0.91), and \textsc{Clarity} (0.75). It also achieved the highest performance in expert-judged \textsc{all} (run-time) with a rate of 0.34, and the second highest performance in expert-judged \textsc{all} (compile) with a rate of 0.26.


The SFT Llama variant demonstrated improvements over its base model, with increases of 0.11 in expert-judged \textsc{Socratic}, and 0.33 in expert-judged \textsc{Selectivity}. A more substantial gain of 0.50 was observed in LLM-judged \textsc{Selectivity} for the SFT variant over its base counterpart. 
Likewise, SFT-Qwen-4B showed improvements across all metrics compared to its base counterpart, with the exception of expert-judged \textsc{No Solution}. The most significant gains for SFT-Qwen-4B were observed in expert-judged \textsc{Socratic}, with a rate of 0.65, and \textsc{Clarity}, with a rate of 0.76.

When compared to the baseline \dcchelp{} responses, all SFT models showed improvement in every metric except for LLM-judged \textsc{Completeness}. The most substantial gains were specifically noted in expert-judged \textsc{Clarity} and in LLM-judged \textsc{No Overhelp}.

\paragraph{\textbf{Expert Inter-Rater Reliability.}}
IRR among the four expert annotators, as measured by Gwet AC1 scores, indicates varying levels of agreement across metrics, with almost perfect agreement for \textsc{Correctness} ($AC1 = 0.84$);
substantial agreement for \textsc{No Overhelp} ($AC1 = 0.72$) and \textsc{Novice Appropriate} ($AC1 = 0.62$); moderate agreement for \textsc{Selectivity} ($AC1 = 0.56$) and \textsc{Completeness} ($AC1 = 0.56$); slight agreement for \textsc{Clarity} ($AC1 = 0.29$); and systematic disagreement for \textsc{Socratic} ($AC1 = -0.12$). 

\paragraph{\textbf{LLM-judge vs Expert IRR}}
IRR across the expert annotations and the LLM-judged annotations, using Gwet AC1 scores, indicate substantial agreement for \textsc{Correctness} ($AC1 = 0.75$), \textsc{No solution} ($AC1 = 0.74$), \textsc{No Overhelp} ($AC1 = 0.71$) and \textsc{Novice Appropriate} ($AC1 = 0.71$); moderate agreement for \textsc{Selectivity} ($AC1 = 0.57$) and \textsc{Clarity} ($AC1 = 0.46$); and slight agreement for \textsc{Socratic} ($AC1 = 0.08$) and \textsc{Completeness} ($AC1 = 0.05$).

%% file: figures/rank_win_rates.tex
\begin{table}[htbp]
\caption{Win-Rate (0-1) of SFT Models Ranked Against Other Models (n = 100) on Compile-/Run-Time Errors. Each Cell: Compile-/Run-Time. Bold = Win-Rate > 0.5}
\label{tab:winrate-compile-runtime}
\resizebox{\linewidth}{!}{%
\begin{tabular}{l|ccccc}
\toprule
\multicolumn{1}{c|}{} & \multicolumn{5}{c}{Loser} \\
\midrule
Winner & \dcchelp{} & GPT-4.1 & Llama-8B & Qwen-32B & Qwen-4B \\
\midrule
SFT-Qwen-4B    & \textbf{0.68} / \textbf{0.78} & 0.34 / 0.36 & \textbf{0.68} / \textbf{0.78} & 0.48 / 0.40 & \textbf{0.70} / \textbf{0.58} \\
SFT-Llama-8B  & \textbf{0.62} / \textbf{0.74} & 0.36 / 0.44 & \textbf{0.74} / \textbf{0.72} & 0.36 / 0.44 & \textbf{0.68} / \textbf{0.74} \\
SFT-Qwen-32B   & \textbf{0.70} / \textbf{0.82} & 0.36 / 0.46 & \textbf{0.78} / \textbf{0.84} & 0.46 / 0.46 & \textbf{0.72} / \textbf{0.80} \\
\bottomrule
\end{tabular}
}
\end{table}

%% file: figures/rank_scores.tex
\begin{table}[htbp]
\caption{Expert Ranking Scores by Model (n = 100)}
\label{tab:expert-ranking}
\resizebox{\linewidth}{!}{
\begin{tabular}{l|rrrr}
\toprule
Model & Mean Rank & 95\% CI & First Place \% & Last Place \% \\
\midrule
GPT-4.1 & 3.09 & 2.68-3.50 & 0.29 & 0.05 \\
SFT-Qwen-32B & 3.60 & 3.20-4.00 & 0.16 & 0.02 \\
Qwen-32B & 3.62 & 3.23-4.01 & 0.16 & 0.05 \\
SFT-Qwen-4B & 4.12 & 3.69-4.55 & 0.15 & 0.06 \\
SFT-Llama-8B & 4.27 & 3.83-4.71 & 0.15 & 0.07 \\
Qwen-4B & 5.50 & 5.12-5.88 & 0.01 & 0.18 \\
\dcchelp{} & 5.67 & 5.24-6.10 & 0.06 & 0.30 \\
Llama-8B & 6.13 & 5.77-6.49 & 0.02 & 0.27 \\
\bottomrule
\end{tabular}
}
\end{table}

%% file: figures/llm_judge/combined_eval.tex
\begin{table*}[htbp]
\caption{Criterion True-Rate (0–1) by Model. Each Cell: Expert/LLM-as-Judge. Gwet’s AC1 is: Expert/Expert--LLM. "all" Columns Report the Rate of Model Responses Satisfying All Criteria. Bold = Highest Rate in Each Column.
}
\label{tab:merged-expert-llm-raw}
\resizebox{1.0\linewidth}{!}{
\begin{tabular}{l|ccc|ccccc|cc}
\toprule
Model & Correctness & Selectivity & Completeness & Clarity & Novice Ap. & No Solution & No Overhelp & Socratic & all (compile) & all (run-time) \\
\midrule
\multicolumn{11}{c}{\textbf{Base models}} \\
\midrule
\dcchelp{} & 0.86 / 0.82 & 0.71 / 0.64 & 0.68 / 0.42 & 0.35 / 0.77 & 0.69 / 0.80 & 0.74 / 0.64 & 0.70 / 0.56 & 0.50 / 0.09 & 0.04 / 0.03 & 0.02 / 0.05 \\
GPT-4.1 & \textbf{0.92} / \textbf{0.97} & \textbf{0.90} / \textbf{0.95} & \textbf{0.87} / 0.45 & 0.75 / \textbf{0.98} & \textbf{0.89} / \textbf{0.99} & 0.82 / \textbf{0.97} & \textbf{0.95} / \textbf{0.97} & \textbf{0.65} / 0.38 & \textbf{0.34} / 0.23 & 0.30 / 0.19 \\
Llama-8B & 0.67 / 0.48 & 0.45 / 0.31 & 0.58 / 0.21 & 0.58 / 0.65 & 0.61 / 0.68 & 0.65 / 0.68 & 0.73 / 0.52 & 0.45 / 0.07 & 0.06 / 0.03 & 0.10 / 0.01 \\
Qwen-32B & \textbf{0.92} / 0.94 & 0.82 / 0.90 & 0.78 / \textbf{0.63} & 0.69 / 0.97 & 0.88 / 0.98 & 0.81 / 0.94 & 0.90 / 0.92 & 0.62 / \textbf{0.44} & 0.22 / \textbf{0.34} & 0.28 / \textbf{0.26} \\
Qwen-4B & 0.74 / 0.70 & 0.61 / 0.65 & 0.59 / 0.23 & 0.59 / 0.81 & 0.65 / 0.81 & 0.86 / 0.89 & 0.86 / 0.83 & 0.50 / 0.15 & 0.12 / 0.07 & 0.10 / 0.03 \\
\midrule
\multicolumn{11}{c}{\textbf{SFT models}} \\
\midrule
Llama-8B & 0.88 / 0.84 & 0.78 / 0.81 & 0.71 / 0.26 & 0.72 / 0.93 & 0.79 / 0.95 & 0.85 / 0.96 & 0.87 / 0.94 & 0.56 / 0.28 & 0.26 / 0.13 & 0.28 / 0.08 \\
Qwen-32B & \textbf{0.92} / 0.94 & 0.85 / 0.93 & 0.79 / 0.38 & 0.75 / 0.97 & 0.85 / 0.98 & \textbf{0.91} / \textbf{0.97} & 0.90 / 0.96 & 0.60 / 0.36 & 0.26 / 0.19 & \textbf{0.34} / 0.14 \\
Qwen-4B & 0.89 / 0.86 & 0.84 / 0.83 & 0.77 / 0.28 & \textbf{0.76} / 0.93 & 0.82 / 0.95 & 0.84 / 0.96 & 0.92 / 0.94 & \textbf{0.65} / 0.28 & 0.24 / 0.13 & 0.30 / 0.09 \\
\cmidrule{1-11}
\textbf{Gwet's AC1} & 0.84 / 0.75 & 0.56 / 0.57 & 0.56 / 0.05 & 0.20 / 0.46 & 0.62 / 0.71 & 0.50 / 0.74 & 0.72 / 0.71 & -0.12 / 0.08 
& & \\
\bottomrule
\end{tabular}
}
\end{table*}

%% file: sections/05_discussion.tex
\newcommand{\correct}[0]{\textsc{Correctness}}
\newcommand{\selective}[0]{\textsc{Selectivity}}
\newcommand{\complete}[0]{\textsc{Completeness}}
\newcommand{\clarity}[0]{\textsc{Clarity}}
\newcommand{\novice}[0]{\textsc{Novice Appropriate}}
\newcommand{\solution}[0]{\textsc{No Solution}}
\newcommand{\overhelp}[0]{\textsc{No Overhelp}}
\newcommand{\socratic}[0]{\textsc{Socratic}}

Our primary contribution is that SFT improves the capability of smaller, open-source language models to generate C programming error explanations for CS1/2 students. 
The smaller SFT models showed significant improvements over their base counterparts in both expert and LLM-judged metrics, including \correct{} (Qwen-4B ${\sim}$20\%; Llama ${\sim}$31\%), \selective{} (Qwen-4B ${\sim}$28\%; Llama ${\sim}$73\%), \complete{} (Qwen-4B ${\sim}$20\%; Llama ${\sim}$22\%), \textsc{Novice Appropriate} (Qwen-4B ${\sim}$17\%; Llama ${\sim}$29\%), and \socratic{} (Qwen-4B ${\sim}$30\%; Llama ${\sim}$24\%). This is corroborated by the expert rankings, where the small SFT models were preferred over their base counterparts in 58\%--74\% of cases, across both compile- and run-time errors, 
and achieved ${\geq}$ 20\% improved mean rank scores. 
Overall, these findings validate the effectiveness of SFT as a strategy for enhancing the pedagogical value of small, open-source models.

Additionally, each of the fine-tuned models exceed the performance of \dcchelp{} on almost all metrics, demonstrating gains of ${\geq}$20\% in \clarity{} and \overhelp{}, and with responses ranked higher in 68\%--82\% of cases. The sole exception was the LLM-judged \complete{} metric, where \dcchelp{} outperformed each of the SFT models. However, the comparatively low LLM-judged rates, along with very weak expert--LLM agreement ($AC1 = 0.05$), suggest a methodological problem with the LLM annotation process for \complete{}, rather than an actual performance shortfall.
Ultimately, all three of the fine-tuned open-source models surpass the baseline standard set by the original \dcchelp{} responses, and so could already be adopted as viable replacements.

When compared to GPT-4.1, SFT-Qwen-4B performs within a margin of 0.10 in all expert-judged metrics, and achieves a mean ranks score  worse by only 1.03. This result is competitive and valuable considering the small parameter size of Qwen-4B. 
The LLM-judged annotations show a larger gap in performance from GPT-4.1 to SFT-Qwen-4B in metrics such as \selective{} and \textsc{Completeness}, though this may be affected by the bias of GPT-4.1 as a judge. Regardless, both LLM and expert annotations demonstrate the competitive performance of SFT-Qwen-4B in \clarity{}, \novice{}, \overhelp{} and \solution{}, when compared to the much larger models. 

While Qwen-32B displays no definitive improvement from the fine-tuning process, both the base and SFT variants come close in performance to GPT-4.1 across expert/LLM-judged metrics. This is a valuable finding as it suggests that Qwen-32B can serve as a suitable replacement for GPT-4.1 when generating the training dataset. This would allow institutions to ensure data privacy during the development and training of specialised open-source models, including models which rely on sensitive data.

\subsection{Limitations} \label{sec:limitations}
Our SFT training dataset was generated using GPT-4.1, therefore our results are inherently bounded by the quality, style and biases of GPT-4.1. This also impacts the interpretability of LLM-judged results which used GPT-4.1 as part of the judge ensemble. The SFT process optimises smaller models to mimic the style and structure of GPT-4.1, consequently using GPT-4.1 as a judge may have favoured responses similar to its own output rather than rewarding pedagogical quality. We sought to mitigate this self-bias by employing an ensemble of diverse models for judgement and validating against a human expert evaluation. 

Similarly, as Qwen-32B and GPT-4.1 were represented in the judging ensemble and as candidate models, the LLM-judged evaluation is vulnerable to self-bias. While we sought to mitigate this by adding a third impartial model (Gemini-2.5-Flash) to the ensemble, and enforcing a unanimous-verdict policy, the markedly higher LLM-judged performance for GPT-4.1 and base Qwen-32B suggests that self-bias persisted. As a result, their relative performance in LLM-judged metrics may be overstated.

The low AC1 scores for \socratic{} and \clarity{}, in both expert and expert--LLM agreement, reduce the reliability of conclusions drawn from these metrics, and they may require refinement.

The training and evaluation dataset included sources from real student errors in CS1/2 courses at a single institution. It may not capture the full spectrum of errors encountered in different contexts, and the risk of over-fitting is present. Further validation before large scale application is required, for example, in other C courses.

Finally, the four experts had no knowledge of the set of models being evaluated, and the order of responses was randomised for each example. The original \dcchelp{} responses likely stood out as the only ones not conforming to the common response structure, potentially affecting their evaluation.

\subsection{Future Work}
Our findings confirm the viability of SFT for model specialisation, and present several promising avenues for future research:

\begin{itemize}
    \item \textbf{Human-preference fine-tuning}: Limited gains in larger models such as Qwen3-32B suggest a potential ceiling for SFT techniques. Exploration into human alignment techniques such as Direct Preference Optimisation (DFO) or Reinforcement Learning from Human Feedback (RLHF) could further refine model behaviour.
    \item \textbf{In-situ evaluation}: In-classroom studies to assess the real-world impact of these fine-tuned models are crucial to produce empirical evidence of learning outcome achievement.
    \item \textbf{On-device model deployment}: Significant gains in the performance of smaller models found in this study highlight the opportunity to explore on-device model deployment, such as in-browser or integrated into a local development environment. Such a deployment would provide a clear solution to ongoing privacy, cost and scalability concerns.
\end{itemize}


%% file: sections/06_conclusion.tex
We investigated the efficacy of SFT with open-source language models on a generated dataset to explain C compiler errors for novice programmers across a range of real student compile- and run-time errors from a large CS1/2 programming cohort. Our findings demonstrate that SFT is a viable and effective strategy for specialising smaller models for pedagogical applications. We observed substantial improvements across a range of quality criteria, most notably clarity, selectivity, and pedagogical appropriateness of generated explanations in smaller models.

We found that SFT is most effective at narrowing the gap between smaller, economical models like Llama-3.1-8B and large, frontier models like GPT-4.1. In all cases, the fine-tuned models outperformed existing tools deployed to the CS1/2 cohort.

This research provides a replicable methodology for creating specialised AI-driven educational tools that are not only cost-effective, but also mitigate the privacy and data security concerns associated with commercial frontier models. By demonstrating a practical pathway to developing high-quality, on-premises or even on-device models, this work empowers educators to leverage advanced AI capability to support student learning.

%% file: sections/07_acknowledgements.tex
\begin{acks}
We would like to thank Google.org who supported this work in the form of a 2023 Google Award for Inclusion Research.
\end{acks}